\documentclass[a4paper,aps,twocolumn,nofootinbib]{revtex4}
\RequirePackage[english]{babel}
\RequirePackage[latin1]{inputenc}
\RequirePackage[T1]{fontenc}
\RequirePackage{mathrsfs}
\RequirePackage{amsmath}
\RequirePackage{amssymb}
\RequirePackage{amsbsy}
\RequirePackage{bm}
\begin{document}
\title{\bf{A simple assessment on inflation}}
\author{Luca Fabbri}
\affiliation{DIME Sez. Metodi e Modelli Matematici, Universit\`{a} di Genova,\\
Piazzale Kennedy Pad.D, 16129 Genova, Italy}
\date{\today}
\begin{abstract}
We consider the torsional completion of gravity with spinor and scalar fields: we show how in this environment conditions of extreme symmetry or specific approximations imply the scalar field to be constant, so that slow-roll takes place and inflation occurs.\\
\textbf{Keywords: Torsion-Gravity, Inflation}\\
\textbf{PACS: 04.20.Cv, 98.80.Cq}
\end{abstract}
\maketitle
\section{Introduction}
In cosmology, inflation still lacks a clear understanding.

First proposed by Alan Guth as an encompassing solutions for a few problems appearing in early stages of the universal evolution \cite{Guth}, inflation is nowadays accepted as a standard element although a mechanism capable of engaging it is yet to be known: one of the most popular of the issues is to have inflation prompted by the presence of a scalar field rolling down a potential hill under the condition that this downward roll occurs very slowly \cite{Linde}.

Slow-roll conditions are met when the scalar potential is nearly flat: the potential flatness is a very specific requirement and therefore insisting on such flatness seems an assumption too artificial to be left without further justification. So far, no such justification has been provided.

But so far, these models of inflation driven by a scalar field have been developed in Einstein gravity, neglecting torsion; if torsion is not neglected, a full theory of torsion-gravity can be employed to give to the scalar field a richer coupling, so that the new interactions might provide the conditions for the slow-roll, and ultimately leading to a conceptually simple understanding of inflation dynamics.

In this paper, we will investigate such opportunity.
\section{Propagating Torsion in Gravity for Spinors and Scalars}
We will follow the basic definitions of \cite{Fabbri:2006xq,Fabbri:2009se, Fabbri:2008rq,Fabbri:2009yc,Fabbri:2011kq,Fabbri:2010rw, Fabbri:2012ag,Fabbri:2013gza}, and here we recall the fundamental notations: in all what follows we require the connection $\Gamma^{\alpha}_{\sigma\nu}$ to have a unique symmetric part and so the torsion $\Gamma^{\alpha}_{[\sigma\nu]}\!=\!Q^{\alpha}_{\sigma\nu}$ is a completely antisymmetric dual  $Q_{\alpha\sigma\nu}\varepsilon^{\alpha\sigma\nu\rho}\!=\!W^{\rho}$ of an axial vector for which $\nabla_{[\alpha}W_{\nu]}\!=\!\partial_{[\alpha}W_{\nu]}\!=\!(\partial W)_{\alpha\nu}$ is defined for later employment \cite{Fabbri:2006xq,Fabbri:2009se, Fabbri:2008rq,Fabbri:2009yc, Fabbri:2011kq, Fabbri:2010rw, Fabbri:2012ag,Fabbri:2013gza}; tensor $R^{\nu}_{\phantom{\nu}\alpha\mu\sigma}$ is the Riemann curvature tensor with contractions $R^{\nu}_{\phantom{\nu}\alpha\nu\sigma}$ and $R^{\nu}_{\phantom{\nu}\alpha\nu\sigma}g^{\alpha\sigma}$ being the Ricci curvature tensor and scalar. And $\phi$ is a generic scalar field which can be interpreted as the Higgs field.

To give the dynamics we assign the Lagrangian \cite{Fabbri:2015rha}
\begin{eqnarray}
\nonumber
&\mathscr{L}\!=\!R\!+\!2\Lambda\!+\!\frac{1}{4}(\partial W)^{2}\!-\!\frac{1}{2}M^{2}W^{2}
\!-\!\frac{1}{2}|\nabla\phi|^{2}+\\
&+\frac{1}{2}\Xi^{2}\phi^{2}W^{2}\!+\!\frac{1}{4}\lambda^{2}\phi^{4}
\!+\!\mathscr{L}_{\mathrm{matter}}
\label{l}
\end{eqnarray}
with $\Lambda$, $M$, $\Xi$ and $\lambda$ being the cosmological constant, the mass of torsion, the torsion coupling to the scalar and the scalar self-coupling, leaving $\mathscr{L}_{\mathrm{matter}}$ as the Lagrangian accounting for all the gauge fields and spinorial matter collectively: by varying (\ref{l}) we obtain, in order, the gravitational field equations as given by the following
\begin{eqnarray}
\nonumber
&R^{\rho\sigma}\!-\!\frac{1}{2}Rg^{\rho\sigma}\!-\!\Lambda g^{\rho\sigma}\!=\!
\frac{1}{2}[M^{2}(W^{\rho}W^{\sigma}\!-\!\frac{1}{2}W^{2}g^{\rho\sigma})+\\
\nonumber
&+\frac{1}{4}(\partial W)^{2}g^{\rho\sigma}
\!-\!(\partial W)^{\sigma\alpha}(\partial W)^{\rho}_{\phantom{\rho}\alpha}+\\
\nonumber
&+\frac{1}{4}\lambda^{2}\phi^{4}g^{\rho\sigma}
\!+\!\Xi^{2}\phi^{2}(\frac{1}{2}W^{2}g^{\rho\sigma}\!-\!W^{\rho}W^{\sigma})+\\
&+\nabla^{\rho}\phi^{\dagger}\nabla^{\sigma}\phi
\!-\!\frac{1}{2}g^{\rho\sigma}|\nabla\phi|^{2}\!+\!T^{\rho\sigma}]
\label{gravity}
\end{eqnarray}
whose contracted form
\begin{eqnarray}
&2(R\!+\!4\Lambda)\!=\!M^{2}W^{2}\!-\!\lambda^{2}\phi^{4}\!-\!\Xi^{2}\phi^{2}W^{2}
\!+\!|\nabla\phi|^{2}\!-\!T
\end{eqnarray}
can be used to write (\ref{gravity}) as
\begin{eqnarray}
\nonumber
&R^{\rho\sigma}\!+\!\Lambda g^{\rho\sigma}\!=\!
\frac{1}{2}[M^{2}W^{\rho}W^{\sigma}+\\
\nonumber
&+\frac{1}{4}(\partial W)^{2}g^{\rho\sigma}
\!-\!(\partial W)^{\sigma\alpha}(\partial W)^{\rho}_{\phantom{\rho}\alpha}
\!-\!\Xi^{2}\phi^{2}W^{\rho}W^{\sigma}-\\
&-\frac{1}{4}\lambda^{2}\phi^{4}g^{\rho\sigma}
\!+\!\nabla^{\rho}\phi^{\dagger}\nabla^{\sigma}\phi
\!+\!(T^{\rho\sigma}\!-\!\frac{1}{2}Tg^{\rho\sigma})]
\label{gravityreduced}
\end{eqnarray}
plus the torsion field equations
\begin{eqnarray}
&\nabla_{\alpha}(\partial W)^{\alpha\nu}\!+\!M^{2}W^{\nu}\!-\!\Xi^{2}\phi^{2}W^{\nu}\!=\!S^{\nu}
\label{torsion}
\end{eqnarray}
with $M^{2}\nabla_{\nu}W^{\nu}\!-\!\Xi^{2}\phi^{2}\nabla_{\nu}W^{\nu}
\!-\!\Xi^{2} W^{\nu}\nabla_{\nu}\phi^{2}\!=\!\nabla_{\nu}S^{\nu}$ and plus the scalar field equations given according to the form
\begin{eqnarray}
&\nabla^{2}\phi\!+\!\lambda^{2}\phi^{2}\phi\!+\!\Xi^{2} W^{2}\phi\!=\!J
\label{scalar}
\end{eqnarray}
in which $T^{\rho\sigma}$ and $S^{\nu}$ with $J$ are the sources due to the matter fields, and whose field equations are not shown.
\section{Homogeneous-Isotropic Universe}
We study what happens in the case of homogeneous and isotropic spacetimes, with identically vanishing spatial curvature: when coordinates $(t,r,\theta,\varphi)$ are used the metric is given in terms of only one function of the cosmological time $A(t)$ written according to the following
\begin{eqnarray}
\nonumber
&g_{\varphi\varphi}\!=\!g_{\theta\theta}(\sin{\theta})^{2}\!=\!g_{rr}r^{2}(\sin{\theta})^{2}
\!=\!-A^{2}r^{2}(\sin{\theta})^{2}\\ 
&g_{tt}\!=\!1
\label{metric}
\end{eqnarray}
with all other components equal to zero while the torsion axial vector has the temporal component $W^{t}$ with all the others vanishing, and the scalar has no restriction apart from having only the cosmological time dependence.

The effects of homogeneity and isotropy on the evolution of the fields has quite an impact when we consider that as $W^{2}\!=\!W^{t}W^{t}g_{tt}\!=\!|W^{t}|^{2}$ and since scalars depend only on the cosmological time then also $W^{t}$ must depend only on the cosmological time, with the consequence that we can deduce also $\partial W\!\equiv\!0$ and therefore (\ref{torsion}) reduce to 
\begin{eqnarray}
&M^{2}W^{\nu}\!-\!\Xi^{2}\phi^{2}W^{\nu}\!=\!S^{\nu}
\label{torsionsymmetric}
\end{eqnarray}
identically; another consequence of isotropy is that the axial-vector source disappears, leaving 
\begin{eqnarray}
&(M^{2}\!-\!\Xi^{2}\phi^{2})W^{\nu}\!=\!0
\end{eqnarray}
as one can easily see. This provides two scenarios: a first is that $W^{\nu}\!=\!0$ giving the known case in which the torsion field vanishes; another is that $M^{2}\!=\!\Xi^{2}\phi^{2}$ therefore giving rise to a constant scalar. The last condition substituted into the remaining field equations (\ref{gravityreduced}, \ref{scalar}) eventually yields
\begin{eqnarray}
&R^{\rho\sigma}\!=\!-\Lambda g^{\rho\sigma}
\!-\!\frac{1}{2}\left|\frac{\lambda}{2}\frac{M^{2}}{\Xi^{2}}\right|^{2}g^{\rho\sigma}
\!+\!\frac{1}{2}(T^{\rho\sigma}\!-\!\frac{1}{2}Tg^{\rho\sigma})
\end{eqnarray}
and
\begin{eqnarray}
&(\lambda^{2}M^{2}\!+\!\Xi^{4}|W^{t}|^{2})M\!=\!J\Xi^{3}
\label{constraint}
\end{eqnarray}
as a constraint fixing the value of torsion in terms of that of the scalar source: notice that in the gravitational field equations an effective cosmological constant term 
\begin{eqnarray}
&\Lambda_{\mathrm{effective}}\!=\!\Lambda
\!+\!\frac{1}{2}\left|\frac{\lambda}{2}\left|\frac{M}{\Xi}\right|^{2}\right|^{2}
\end{eqnarray}
is generated. Notice that it is positive and it can be large if the ration given by $M/\Xi$ is large as we may have in the case the torsion-scalar coupling constant were small.

By computing the energy tensor for the scalar field in the pair of cases $W^{\nu}\!=\!0$ and 
$M^{2}\!=\!\Xi^{2}\phi^{2}$ we see that the former would consist in the potential energy plus the kinetic energy while the latter would consist only in the potential energy, and since the kinetic energy is always positive, the latter condition is energetically favoured. 

The latter condition is not only an additional possibility, but also the case the scalar would tend to favour thus showing how a slow-roll could be therefore justified.
\section{Effective Approximation}
In what we have done thus far, the condition of perfect homogeneity and isotropy have been essential; but on the other hand, although conditions of perfect symmetry may be reasonable in the early universe, nevertheless this may not be the case in general and it would be preferable to have a way to engage the above mechanism in conditions of less symmetry: this is addressed by acknowledging that to engage such a mechanism in the most general case that is possible, we must first remove axial-vector spin sources and then find the most general symmetry for which the kinetic torsion term in field equations (\ref{torsion}) disappears.

For the first task we may notice that although the lack of isotropy may allow the spin of the single spinor to be different from zero, nevertheless we would still have at a local level that the orientations of the spin of all spinors would average away so that the axial-vector source would still disappear; for the second task we have to see under what conditions on the metric is the relationship
\begin{eqnarray}
&\partial_{\alpha}(\sqrt{|g|}g^{\alpha\rho}g^{\nu\mu}\partial_{[\rho}W_{\mu]})\!=\!0
\end{eqnarray}
valid for any torsion, and because this amounts to look for $4$ constraints over a field with $10$ degrees of freedom then we have that in general a solution is possible.

However, there is also the possibility to approach the problem from an entirely complementary perspective, by noticing that although in this model torsion is in general a propagating field, its mass might be large enough as to allow effective approximations giving rise to
\begin{eqnarray}
&(M^{2}\!-\!\Xi^{2}\phi^{2})W^{\nu}\!\approx\!0
\end{eqnarray}
already in early stages of universal evolution, and regardless any condition of whatsoever symmetry for the metric.

Even more importantly, the energy of the scalar would be left unchanged and therefore the condition providing a constant scalar field would remain the most favourable.

Therefore slow-roll would even be better justified.
\section{Comments}
In order to conclude the discussion, it is necessary to comment about the eventuality that inflation may last forever, specifying that this is not the case in the present model; to convince oneself of this, it is enough to consider the scalar field equation in the form of the constraint given by the expression (\ref{constraint}): after inflation has stretched the universe, larger volumes imply smaller material densities with the consequence that the scalar source $J$ would vanish and thus (\ref{constraint}) would no longer possess a solution.

The scalar field equation with a scalar source ensures that there be some lower limit in the matter density and therefore an upper limit of the volume of the universe undergoing inflation; when this limit is reached, a different dynamical behaviour of the universal evolution occurs.

Physically, we may see this limit as the scale at which the energetically favourable solution would become that of vanishing torsion and no longer that of constant scalar field, the discussed mechanism would no longer work and the inflation as due to this model would stop.
\section{Conclusion}
In this paper, we showed that if propagating torsion in gravity is coupled to spinor and scalar fields it gives rise to the possibility that homogeneity and isotropy imply a constant scalar field, justifying the slow-roll mechanism leading to inflation in what we believe to be one of the simplest ways; this mechanism recalls that of super-cooling, where liquids can be brought below their freezing temperature without actually freezing so long as their purity ensures a symmetry in the density distribution.

We have also discussed that despite being propagating torsion has a mass term, and if large enough this might provide the conditions of effective approximations which would allow, even with no global symmetries, to neglect the kinetic torsion term, so to give a constant scalar field, then an exponential expansion, and so inflation; in fact, this situation would be even more physical because in it inflation would not always be present, and only when the ratio between temperature and torsion mass allows the effective approximation inflation would engage.

Opportunity for future work may be sought in rendering this mechanism more detailed, by specifying the exact dynamics of the matter distributions.

\end{document}